\documentstyle[12pt,aasms4,epsfig]{article}

\def\amin{\ifmmode^{\prime}\else$^{\prime}$\fi}
\def\asec{\ifmmode^{\prime\prime}\else$^{\prime\prime}$\fi}

\def\simgt{\lower.5ex\hbox{$\; \buildrel > \over \sim \;$}}
\def\simlt{\lower.5ex\hbox{$\; \buildrel < \over \sim \;$}}

\lefthead{Mukherjee, et al. }
\righthead{}

\begin{document}

\title{Multiwavelength Examination of the COS--B Field 2CG 075+00
 Yields a Blazar Identification for 3EG J2016+3657}

\author{R. Mukherjee} 
\affil{Dept. of Physics \& Astronomy, Barnard College \& Columbia University, New York, NY 10027}

\author {E. V. Gotthelf \& J. Halpern}
\affil{Dept. of Physics \& Astronomy, Columbia University, New York, NY 10027}

\author{M. Tavani}
\affil{IFCTR-CNR, via Bassini 15, I-20133 Milano, Italy and Columbia 
Astrophysics Lab, New York, NY 10027}

\bigskip
\bigskip

\centerline{Accepted by {\sl The Astrophysical Journal}}

\begin{abstract}

We present a high-energy study of the intriguing COS--B $\gamma$-ray 
field, 2CG 075+00, in order to search for possible 
counterparts. New EGRET data show that the COS--B emission probably 
corresponds to two localized $\gamma$-ray sources, 
3EG J2016+3657 and 3EG J2021+3716. Spectral fits to these EGRET sources, 
assuming 
a power-law model, yield photon indices of $\sim 2$ for each object. 
We examine archival ROSAT and ASCA X-ray data which overlap 
both EGRET error boxes, and find several point sources in the 
region to a flux limit of approximately $6.5\times 10^{-13}$ erg cm$^{-2}$
s$^{-1}$. We conclude that the 
most probable candidate for 3EG J2016+3657 is the compact, variable, 
flat-spectrum radio and millimeter source B2013+370 (G74.87+1.22) 
which has blazar-like properties. 
The other source, 3EG J2021+3716, remains unidentified. 

\end{abstract}


\section{Introduction}

Since the first surveys of the $\gamma$-ray sky with the COS--B
satellite the nature of most $\gamma$-ray sources remain a mystery, as
few of these sources have firmly established counterparts at any other
waveband (Swanenburg et al. 1981).
With the launch of the {\sl Compton Gamma Ray Observatory} (CGRO) in 1991,
improved surveys at relatively higher angular resolution 
in the $\gamma$-ray band was made possible. In particular, a systematic 
survey of the $\gamma$-ray sky, including the COS--B source regions was 
carried out with the on-board EGRET (Energetic Gamma-ray Experiment Telescope) 
instrument at energies above 100 MeV. EGRET has so far detected 271 
sources of high energy $\gamma$-rays in the third EGRET (3EG) catalog (Hartman 
et al. 1999), of which 169 remain unidentified (74 of these are at $\vert
b\vert<10^\circ$), with no convincing counterparts at other wavelengths. 
Surprisingly, only two of the unidentified COS--B
sources have been subsequently associated with EGRET sources, and both
are pulsars, namely Geminga (Bertsch et al. 1992), and 2CG 342-02 (PSR
B1706-44) (Thompson et al.  1992). A third source, PSR B1046-58, could 
possibly be the candidate for identification of 2CG 288-00 
(Kaspi et al. 2000). 

To date, the only sources of high energy $\gamma$-rays convincingly
identified are the blazars at high galactic latitudes and the pulsars at low 
latitudes. Studies of individual unidentified EGRET $\gamma$-ray source fields 
have been inconclusive in locating the origin of their emission. Comprehensive 
surveys of the field associated with these sources have met with limited
success (see Mukherjee, Thompson \& Grenier 1997 for a review). Several 
researchers have noted that the unidentified EGRET
sources in the Galactic plane lie in proximity to star formation sites
and supernova remnants (Yadigaroglu \& Romani 1997, Sturner \& Dermer
1995, Esposito et al. 1996), while others report a correlation with OB
associations and massive stars (Kaaret \& Cottam 1996; Kaul \& Mitra 1997; 
Romero et al. 1999).  
Efforts to identify the $\gamma$-ray sources at other wavelengths include 
systematic multifrequency radio observations (e.g. \"Ozel et al. 1988)
and X-ray imaging studies (e.g., Brazier et al. 1996; 1998;
Roberts \& Romani 1998; Mirabal et al. 2000).

In this article we re-visit the region containing the unidentified
COS--B source 2CG 075+00, located in the Cygnus region, for which a 
significant amount of archival
$\gamma$-ray (EGRET) and X-ray (ASCA \& ROSAT) data have 
accumulated. Previous attempts to locate the origin of the high energy
emission were fraught with frustration, as the position
associated with 2CG 075+00 in the second EGRET catalog (Thompson et al. 1995), 
2EG J2019+3719, has shifted in the 3EG catalog, and has split into 
two discrete sources, 3EG J2016+3657 and 3EG J2021+3716. Fortunately, both
revised EGRET error boxes have overlapping archival ASCA and ROSAT
observations. We now re-analyze the EGRET data along with the
corresponding X-ray fields using the refined $\gamma$-ray positions
from the later catalog.

\section{The $\gamma$-ray Observations}

2CG~075+00, first observed by COS--B, is located in the Galactic plane. 
The revised COS--B position for the source is $l=76.1^\circ$, $b=0.5^\circ,$ 
with an error radius of $\sim 1.0^\circ$ (Pollock et al. 1985). In the 
Second COS--B Catalog, Swanenburg et al. (1981) note that the 
source structure could possibly be interpreted as containing 
extended features. 
The integrated $\gamma$-ray flux from the source at energies greater than 100 MeV 
was found to be $1.3\times 10^{-6}$ photons cm$^{-2}$ s$^{-1}$, although no 
spectral information was available (Swanenburg, et al. 1981). 

Since its launch in 1991, EGRET has observed the error circle of 2CG~075+00 
several times. EGRET is sensitive to $\gamma$-rays in the energy 
range from 30 MeV to 30 GeV, with a point source sensitivity of 
$\sim 3\times 10^{-7}$ ph cm$^{-2}$ s$^{-1}$ ($> 100$ MeV). It has an effective 
area of $1.5\times 10^3$ cm$^2$ in the energy range 0.2 -- 1 GeV, 
decreasing to about one-half the on-axis value at $18^\circ$ off-axis and to 
one-sixth at $30^\circ$. The nominal angular resolution is 
$0.6^\circ$ at 500 MeV, improving to $0.4^\circ$ at 3 GeV. Details of the 
instrument preflight and in-flight calibration are described elsewhere 
(Thompson et al. 1993; Esposito et al. 1999). 

In the second EGRET (2EG) catalog, 2CG 075+00 was weakly detected as
2EG J2019+3719. Reanalysis of the region using a larger data set for
the 3EG revealed two sources, 3EG J2016+3657 and 
3EG J2021+3716, with error radii of $33^\prime$ and $18^\prime$, 
respectively, at the 95 \% contour (Hartman et al. 1999). 
In the summed Phase 1 through Cycle 4 EGRET observations, 
3EG J2016+3657 and 3EG J2021+3716 were detected at the highest significances 
of $6.4 \sigma $ and $10.3\sigma$, respectively, and these observations were 
used for the position determination. Figure 1 shows the EGRET source positions 
superimposed on the ROSAT X-ray image that is described in \S 3. 
3EG J2021+3716 is completely within the 
error circle of the COS--B source 2CG 075+00, also shown in Figure 1. 

An analysis of 4.5 years of EGRET data based on photons with energies greater 
than 1 GeV gives a slightly different result for source positions in the 
Cygnus region (GeV catalog: Lamb \& Macomb 1997). The GeV catalog 
lists one source with an error circle overlapping that of 2CG 075+00, namely, 
GEV J2020+3658, at $l=75.29^\circ$, $b=0.24^\circ$ (see Fig. 1). 

 We examined the flux history of 3EG J2016+3657 and 3EG J2021+3716 to search 
for variability, such as
is characteristic of high latitude EGRET blazars. The light curve for
these sources are displayed in Fig. 2 using data from the 3EG catalog 
(Hartman et al. 1999) for the Phase 1 through Cycle 4 
observations (1991-1995). Two additional observations were made in Cycle 6 in 
viewing period (VP) 601.1 (1996 October) and VP 623.5 (1997 July). These 
data were analyzed using the standard EGRET data processing technique, 
as described in Mattox et al. (1996) and Hartman et al. (1999), and are 
included in Fig. 2. 
The horizontal bars on the individual data points denote
the extent of the VP for that observation.  Fluxes have been plotted
for all detections greater than $2\sigma$.  For detections below
$2\sigma$, upper limits at the 95\% confidence level are shown.  
A $\chi^2$ analysis can be used to calculate a 
variability index according to that defined by McLaughlin et al. 
(1996).  Although somewhat arbitrary, the quantity $V$ can be used 
to judge the strength of the evidence of flux variability. Following the 
classification used in McLaughlin et al. (1996), we use $V<0.5$ to indicate 
non-variability, and $V\geq 1$ to indicate variability. In this 
case, we obtain $V=1.1$ for 3EG J2016+3657 and $V=1.57$ for 
3EG J2021+3716. We note, however, that 3EG J2016+3657 and 3EG J2021+3716 are 
in a confused region (as indicated in the 3EG Catalog), and the variability 
numbers could be an artifact of the analysis, rather than being intrinsic to 
the sources. 
In comparison to the EGRET blazars, 3EG J2016+3657 and 3EG J2021+3716 
have variability indices similar to a large fraction of the 
blazars detected by EGRET (Mukherjee et al. 1997). 

The background-subtracted $\gamma$-ray spectra of 3EG J2016+3657 and 3EG
J2021+3716 were determined by dividing the EGRET energy band of 30 MeV
-- 10 GeV into 4 bins for 3EG 2016+3657 and 10 bins for 3EG 2021+3716, 
and estimating the number of source photons in
each interval, following the EGRET spectral analysis technique of
Nolan et al. (1993). The data were fitted to a single power law of the
form $F(E) \propto (E)^{-\Gamma}$ photon cm$^{-2}$ s$^{-1}$
MeV$^{-1}$, where $F(E)$ is the flux, $E$ is the energy, $\Gamma$ is
the photon spectral index.  The photon spectral indices of 3EG
J2016+3657 and 3EG J2021+3716 were found to be, respectively, 
$1.99\pm 0.20$ and $1.86\pm 0.10$. 

\section{The X-ray Observations}

The error boxes of both 3EG sources are covered by archival X-ray
imaging observations acquired with the ROSAT and ASCA 
observatories. Two adjacent observations with each
observatory fall nicely on the two 3EG error boxes.  Historically, these 
fields have been studied in X-rays both because of the existence 
of the 2CG source, as well as due to several known X-ray sources 
in the region. The ASCA observations herein were a part of a 
program to study unidentified sources in the Galactic plane (Tavani et al. 
2000). 

Archival data for the region were available for the ROSAT 
Position Sensitive Proportional Counter (PSPC), the ROSAT High Resolution 
Imager (HRI) and the ASCA Gas
Imaging Spectrometer (GIS) which allow complementary broad-band X-ray
data in the 0.2 -- 2.0 keV (PSPC) and 1 -- 10 keV (GIS) 
range with arcmin spatial resolution and
moderate energy resolution. The PSPC $1^\circ$ radius field-of-view
is about twice that of the GIS. The ROSAT HRI observations took place on 
1994 November 12 -- 13, with a total exposure time of 43 ks. The ROSAT 
PSPC observations were on 1991 November 22 -- 30, 1992 April 28 -- 30, 
1993 October 24 -- 25 and 1994 June 5, with a 
total exposure time of $\sim 12$ ks. The ASCA observations took place 
on 1995 May 29 -- 31, 1995 October 14 -- 15, and 1996 February 2. 
All data were obtained from the HEASARC
archive at Goddard Space Flight Center and edited using the latest
standard processing for each mission. 

We created ROSAT (Fig. 1) and ASCA (Fig. 3) images of the region 
containing 3EG J2016+3657 and 3EG J2021+3716 by co-adding exposure corrected 
sky maps from each mission. These images are centered on the
position of the earlier Second Catalog source, 2EG 2019+3719. However,
the PSPC image size is large enough to include the 95~\% error
contours of both the 3EG sources. 
Note that the ASCA images are not centered on the EGRET 
positions, and only part of the error contour of 3EG
2016+3657 is covered by the ASCA observation. In both Fig. 1 and Fig. 3 
we show the error circles of 2CG 075+00 with dashed lines. 
Also shown is the error circle of the 
GeV source GeV J2020+2658 (Lamb \& Macomb 1997). 
 
To search for a possible X-ray counterpart to the $\gamma$-ray
sources, we examined the ROSAT point sources enclosed by the 95~\% 
contours of the two 3EG sources. The detected positions are 
numbered in the image (Fig. 1) 
and are tabulated in Table 1. There are 9 bright 
sources in the ROSAT image of 3EG J2016+3657, but none that are significant 
within the error circle of  3EG J2021+3716. Note that Source 3 (indicated with 
an arrow) is very faint,
and barely resolved in the ROSAT PSPC image. Its position is determined from the
ROSAT HRI image shown in \S 4. 
There are at least 5 X-ray sources in the error circle 
of 2CG 075+00, which reached a minimum detectable intrinsic flux of 
$6.5 \times 10^{-13}$ erg cm$^{-2}$ s$^{-1}$ in the 0.1 -- 2.4 keV band, 
assuming a power-law photon spectral index of 2.0, and a Galactic 
column absorption of $1\times 10^{22}$ cm$^{-2}$.  

The ASCA images were, similarly, searched for corresponding X-ray 
counterparts. In deriving the ASCA positions, we were able to use the 
ROSAT point sources seen in the ASCA images to improve the astrometry for the
ASCA sources to $10^{\prime\prime}$ by registering the ASCA images
using the overlapping ROSAT sources. There are no significant 
point sources in the ASCA image within the 95~\% contour of 3EG
J2021+3716. The ASCA image of 3EG J2016+3657 reveals 5 point sources which 
are indicated with numbers in Fig. 3. Source numbers 1, 2, 3, 4 and 5
correspond to ROSAT sources of the same numbers in Fig. 1. 

To measure the ASCA and ROSAT source count rates we extracted photons using a
$2^{\prime}$ radius aperture and estimated the background contribution
using a large annulus away from the other source following the method
described in Gotthelf \& Kaspi (1998).  For ASCA, we define hardness ratio 
as 
$${{S(<2\ {\rm keV}) - S(>2\ {\rm keV})}\over{ S(<2\ {\rm keV}) + S(>2\ {\rm keV})}}.$$  
The ASCA and ROSAT sources are listed in Table 1 along with their background
subtracted count rates, detection significances, and hardness ratios. 
We found no other point sources in the ASCA image at the level of 
$5\sigma$ or higher, other than those listed in Table 1. 

We have searched for counterparts of the X-ray sources in the ROSAT and ASCA 
images. Several of the sources have optical identifications and are listed 
in Table 1. Source 1 is a known supernova remnant, CTB 87. Two of the sources 
can be identified with radio sources. Notes on the individual sources in 
Table 1 are given in the following section. 

\section{Notes on Individual Sources}

{\sl CTB 87}: Source number 1 in Table 1 is coincident with 
SNR CTB 87 (G74.9+1.2), an extended source with a flat radio spectrum. 
G74.9+1.2 is a filled-center SNR in the radio with high polarization and a 
high frequency turnover. Its HI absorption indicates a distance of 12 kpc. 
It has a relatively flat spectrum in the radio with a 
spectral index of $0.26\pm 0.2$ below 11 GHz, beyond which a transition to a 
steeper spectral index occurs with the spectrum steepening to an index $> 1$. 
(Morsi \& Reich 1987; Salter et al. 1989). It is generally believed that 
filled-center SNRs are remnants in which a central object is responsible 
for the relativistic electrons, whose synchrotron emission is detected at 
radio frequencies (Koyama et al. 1997). The flat radio spectrum 
is believed to be due to the central pulsar which injects particles into the 
nebula (Reynolds \& Chevalier 1984). 

\medskip

{\sl B2013+370}: 
The hard X-ray source marked `3' in the ASCA image (Fig. 3), that is barely 
resolved in the ROSAT PSPC image (Fig. 1), is consistent with the radio 
source B2013+370 (G74.87+1.22). 
B2013+370 is a well-studied compact, flat-spectrum radio source that was 
first noticed during the study of the SNR CTB 87 (Duin et al. 1975). 
It is located approximately $7^\prime$ west of the brightness peak of CTB 87. 
Wilson (1980) first noted the 
association of the radio source with an extended or possibly double 
X-ray source observed in the 
0.15 -- 3.0 keV band with the Imaging Proportional Counter (IPC) on the 
{\sl Einstein} Observatory. The IPC source, 1E 2013.7+3701, was situated 
roughly between the locations of our sources 2 and 3. The 
ROSAT HRI image (Fig. 4), however, shows that the Einstein 
source is clearly resolved into two point sources, corresponding to 
sources 2 and 3. 

Due to its proximity to the SNR, the possibility that B2013+370 and CTB 87 
are related cannot be excluded. However, such an association is unlikely based 
on the requirements of an unprecedented velocity for the radio source if it 
were to be associated with the nearby SNR (see for example the arguments 
presented in Shaffer et al. 1978). 

In fact, B2013+370 has all the characteristics of 
a compact, extragalactic, non-thermal radio source and is typical of the 
many extragalctic sources seen by EGRET. 
Duin et al. (1975) report a radio spectral index of $\alpha = -0.2$ at high 
frequencies (above $\sim 7500$ MHz), 
and a low frequency spectral index of about $\alpha=+0.4\pm0.06$. 
The source exhibits a lack of recombination line emission, the presence of 
linear polarization and a spectrum consistent with a non-thermal source 
showing synchrotron self absorption below 8 GHz, as expected for a magnetic 
field of $<0.1$ Gauss (Duin et al. 1975). VLBI 
measurements indicate an angular extent of $<0.001^{\prime\prime}$ at 8 GHz 
and $> 0.009^{\prime\prime}$ at 0.8 GHz (Weiler \& Shaver 1978). 
From its radio properties, B2013+370 is very likely to be a flat-spectrum radio 
quasar or a BL Lac object. It has a 5 GHz flux of about 2 Jy (Duin et al. 
1975), typical of many blazars seen by EGRET. 
In addition, B2013+370 is detected at 90 GHz and 142 GHz with the IRAM 
30 m telescope and exhibits variability at these 
wavelengths, a specific property of EGRET 
blazars (Bloom et al. 1999, Mattox et al. 1997). 
This is demonstrated in Fig. 5 which shows the light curves of 
B2013+370 at 90 GHz and 142 GHz, measured between 1993 and 1995 
(Reuter et al. 1997). 

\medskip
{\sl HD 228766}: This is a Wolf-Rayet star, also known as SAO 69765 
(Hog et al. 1998), and 
is the counterpart to source number 6 in Table 1. It has a $B$ magnitude of 
9.72 and a $V$ magnitude of 9.22 Its spectral type is O5.5f. 

\medskip
{\sl HD 193077}: Source number 7 in Table 1 is a bright ($B=8.34$, $V=8.06$) 
Wolf-Rayet star, HD 193077 (Perryman et al. 1997), also known as SAO 69755. The 
star corresponds to WR 138 in the Wolf-Rayet catalog (van der Hucht et
al. 1981). Its a star of the WN sequence (subtype WN5+OB), with its spectra 
dominated by broad emission lines of helium and nitrogen (Lepine \& Moffat 1999). 

\medskip
{\sl CCDM J20215+3758A}: Together with CCDM J20215+3758B, this corresponds to 
a double star system and is the counterpart to Source 12 in Table 1. The 
USNO-A2.0 catalog gives the magnitudes of the two stars as follows: 
CCDM J20215+3758A: $R=11.6$, $B=12.1$, and 
CCDM J20215+3758B: $R=11.9$, $B=13.5$. 

\medskip
{\sl HD 229153}: This is the counterpart to source 13 in Table 1 (Hog et
al. 1998). It is a bright star ($B=10.09$, $V=9.14$) of spectral type BOIab. 

\medskip
We also find three of the other X-ray sources to be coincident with 
bright stars in the USNO-A2.0 catalog. Source 8 is possibly a star 
($B=12.4$, $R=11.3$) with coordinates (J2000) 20 16 37.55, +37 05 55.0, and 
Source 9 is most likely a star of approximately 12th magnitude at (J2000) 
20 17 35.86, +36 38 02.3. Source 14 probably corresponds to a star 
($B=11.7$, $R=10.8$) with coordinates (J2000) 20 19 44.16, +37 35 26.7. 
In addition, we find two point sources in the 
ROSAT HRI image (Fig. 4), not listed in Table 1, to be coincident with 
bright stars. These are marked in the figure as: HD 228600 
at (J2000) 20 15 30.8,  +37 20 03.1 ($B=10.5$, $V=10.1$), 
and a star ($B=15.7$, $R=14.2$) at (J2000) 20 16 49.0, +36 57 48. 

We do not find any likely counterparts in the literature 
to the other remaining sources in Table 1. 

\subsection{Optical Observations} 

We obtained CCD images in the $R$ band using the 2.4m telescope of the
MDM Observatory on 2000 April 24, and in the $I$ band on March 18 using the 
MDM 1.3m. The regions covered were $2^{\prime} \times 2^{\prime}$ on the 
2.4m, and $8^{\prime} \times 8^{\prime}$ on the 1.3m. In seeing of 
$0.^{\prime\prime}75$, an
optical object of $R = 21.6 \pm 0.2$ is detected at (J2000)
20 15 28.76, +37 10 59.9 in the USNO--A2.0 reference system
(Monet et al. 1996),
with an uncertainty of $0.^{\prime\prime}3$.
This coincides with the NVSS position (Table 2) of the blazar B2013+370,
identified with X-ray source 3, as shown in Figure 6.
This object was also detected optically by Geldzahler et al. (1984),
who found $I = 19.5 \pm 0.5$, while
also noting that it appeared extended.  Our $R$-band image shows that 
the object closest to the radio position is unresolved, while a fainter 
object $1.^{\prime\prime}7$ to the northwest was almost certainly 
responsible for the extended appearance in the Geldzahler et al. (1984) 
image. 
Assuming standard conversions of $N_{\rm H}$ to extinction, the absorption
in the $R$ band is $\approx 5.1$ mag to an extragalactic object,
making its intrinsic magnitude $R = 16.5$. Our $I$-band images have 
inferior seeing, but they also detect the blended pair at the radio position, 
as well as optical objects at the positions of X-ray sources 2 and 4. 
However, in the absence of further information such as radio detection or
optical spectroscopy,
the severe crowding and Galactic extinction makes the identification of 
these additional X-ray sources 
by positional coincidence alone dangerous, even from HRI positions.

\subsection{Other Radio Sources}

We have searched the NRAO/VLA Sky Survey (NVSS) catalog (Condon et al. 1998) 
of 1.4 GHz radio sources for other possible counterparts to the X-ray and 
$\gamma$-ray sources. A search within the error box of 3EG J2016+3657 revealed 
126 radio sources. Of these, only 5 had integrated radio fluxes $> 0.3$ Jy 
and are shown in Fig. 7 and Table 2. The two 
brightest of these are positionally coincident with the ROSAT and ASCA 
sources 1 (CTB 87) and 3 (B2013+370). The two sources clustered near CTB 87 
are actually extended regions within the supernova remnant. 
None of the other NVSS sources 
match the positions of the X-ray sources in the ASCA and ROSAT images. 

There are two bright $(>0.3$ Jy) radio sources within the error circle of 
3EG J2021+3716 (see Fig. 7, Table 2). Both of these were also detected by IRAS, 
and were shown to be H {\tt II} regions from 
their radio recombination lines (Lockman 1989). 

A search of the Westerbork Northern Sky Survey (WENSS) catalog 
(Rengelink et al. 1997) of 92 cm (325 MHz) radio sources yielded similar 
results. This survey has a limiting flux density of about 18 mJy. A search 
within the error box of 3EG J2016+3657 yielded 21 radio sources, of which 
only 2 matched the positions of the X-ray sources. These are 
(a) WNB2014.1+3703 positionally 
coincident with ROSAT and ASCA source numbers 1, or CTB 87 and 
(b) WNB2013.6+3701 matching the ASCA source 3, or B2013+370. 
A search within the error circle of 3EG J2021+3716 yielded a pair of 
very bright radio source, WNB2019.7+3718B and  WNB2019.7+3718C, which 
correspond to the H{\tt II} regions also seen in the NVSS data (see Table 2). 

\section{Discussion}

Our study of archival X-ray (ASCA and ROSAT) data yields several faint 
sources within the error boxes of the two 3EG sources. The region contains 
bright stars, the 
SNR CTB 87, a compact radio source, and HII regions. 

The presence of the SNR CTB 87 (G74.9+1.2) in the field of 3EG J2016+3657 is 
potentially quite important in light of the $\gamma$-ray source/SNR
associations noticed in previous investigations (Sturner \& Dermer 1995; 
Esposito et al. 1996). 
Gamma-ray production from SNRs, Wolf-Rayet stars and OB associations is 
expected in several theoretical models. This subject was extensively 
investigated  in the past for COS--B sources (Montmerle 1979; 
V\"olk \& Forman 1982), and recently for EGRET (Romero et al. 1999; 
Esposito et al. 1996; Kaaret \& Cottam 1996). If the $\gamma$-ray emission 
were from a young pulsar associated with the SNR, a conclusive way to prove 
this would be to find pulsations. This is unfortunately not possible for 
such a weak X-ray source. 

However, the energetics of the CTB 87 remnant are far from adequate to
produce a source detectable by EGRET at the inferred distance of 12 kpc.
Wilson (1980) argued convincingly that the X-ray luminosity of CTB 87,
which is 100 times less than that of the Crab, implies that the spin-down
power of the embedded pulsar must be correspondingly less,
$I\Omega\dot \Omega \approx 1 \times 10^{36}\ (d/12\ {\rm kpc})$~ergs~s$^{-1}$.
Even assuming that 100\% of this power is emitted in the EGRET energy
band, the resulting flux of $6 \times 10^{-11}$ ergs~cm$^{-2}$~s$^{-1}$
is 20 times less that the EGRET measured average flux of
3EG~J2016+3657.  This argument is insensitive to distance as long as the
X-ray synchrotron nebula is considered a calorimeter of the present
pulsar power.  If, instead, CTB 87 hosts a Geminga-like pulsar whose
energy is no longer trapped by the nebula, and is maximally efficient
in the production of $\gamma$-rays, then we would expect a spin-down
power of only $\sim 3 \times 10^{34}$~ergs~s$^{-1}$.  Such a pulsar
is inadequate to explain the flux of 3EG~J2016+3657 unless it were
at $d < 500$~pc, which is certainly ruled out by the H~I and X-ray
measured column density to the SNR.  
We believe that the SNR CTB 87 is an interesting Crab-like remnant,   
but is too weak and too far away to be a good
candidate for the EGRET source 3EG J2016+3657. 

Similarly, neither the bright stars in the X-ray images nor the H {\tt II} 
regions are likely to be responsible for the EGRET source. 

We believe that the most likely candidate for 3EG J2016+3657 is the radio 
source B2013+370, described in \S4. Based on its radio properties B2013+370 
is very likely to be a blazar, similar to the others seen by EGRET. 
Mattox et al. (1997) find that only the brightest radio-flat AGN can be 
identified with EGRET sources with any level of confidence, and demonstrate 
that there is a high degree of correlation between $\gamma$-ray and radio fluxes 
of EGRET blazars. EGRET has detected more than 
65 active galactic nuclei (AGN) (Hartman et al. 1999), almost all of which 
can be classified as blazars. 
The blazars seen by EGRET all share the common characteristic that they are 
radio-loud, flat spectrum sources, with radio spectral indices 
$0.6>\alpha>-0.6$ (von Montigny et al. 1995). Several of these blazars 
exhibit superluminal motion of components resolved with VLBI (e. g., 
3C 279, 3C 273, PKS 0528+134). The blazar class of AGN 
includes highly polarized quasars, BL Lac objects, or optically violent 
variable (OVV) quasars. The sources are characterized by one or more 
properties of this source class, namely, a flat radio spectrum, a continuum 
spectrum that is non-thermal, optical polarization and strong variability. 
For most EGRET blazars, the $\gamma$-ray luminosity dominates that other 
wavebands. 

The probability that the blazar B2013+370 is the correct identification for the 
EGRET source 3EG J2016+3657 can be estimated by following the calculations 
presented by Mattox et al. (1997). The a priori probability that EGRET 
will detect a random flat-spectrum blazar with a 5 GHz flux of 2 Jy is 
5.8\% (Mattox et al. 1997). However, since in this case there is an EGRET 
source at this location, the conditional (a posteriori) probability must be 
used, which takes into account the fact that a gamma-ray source has already 
been detected. The a posteriori probability that a blazar of the type that 
we see in the error circle of an EGRET source is in fact the correct 
identification is about 98.8\%. 
The factors that enter into this calculation are the radio flux (2 Jy)
and spectral index (+0.3), the 95\% error radius of the EGRET source 
(0.55$^\circ$), the distance of the radio source from the center of the EGRET 
circle (0.27$^\circ$), and the mean distance between radio sources which are at
least as strong and at least as flat as this one 
($\sim 16.7^\circ$.)  

Fig. 8 shows the spectral energy distribution of B2013+370, 
assuming that it is the EGRET source 3EG J2016+3657. The figure shows 
the relative amounts of energy detected in equal logarithmic 
frequency ranges. The radio fluxes were 
obtained from the NRAO/VLA Sky Survey, the Westerbork Northern 
Sky Survey (see \S 4.2 ), and from a compilation of radio fluxes 
in Weiler and Shaver (1978). The fluxes at mm wavelengths were obtained 
from Reynolds et al. (1997), taken with the IRAM 30 m telescope. 
The optical point was measured by us at the MDM Observatory and has been 
corrected for extinction (\S 4.1). 
The estimated unabsorbed X-ray flux for source 3 (=B2013+370) in the 
Asca band of 1 -- 10 keV is $(6\pm1) \times 10^{-13}$ erg cm$^{-2}$ s$^{-1}$, 
and in the ROSAT band of 0.1 -- 2.4 keV is 
$(1.9\pm 0.2)\times 10^{-12}$ erg cm$^{-2}$ s$^{-1}$. 
For the flux estimations we have assumed 
$N_H=10^{22}$ cm$^{-2}$, and a power-law photon index of 2. 
The data in Fig. 8 are not contemporaneous. 
The EGRET spectrum was derived as explained in \S 2.  

The broad band spectrum shown in Fig. 8 is that of a typical EGRET blazar, 
dominated by the power output in $\gamma$-rays.  
The spectrum shows the characteristic features of a blazar, with 
a synchrotron peak at lower energies, and an inverse Compton peak 
at higher energies. The high 
$\gamma$-ray luminosity of a blazar suggests that the emission is likely 
to be beamed, and therefore Doppler-boosted, along the line of sight. 
In this scenario, synchrotron radiation from 
high-energy electrons in a relativistically outflowing jet are 
responsible for the radio to UV continuum. The high-energy photons 
come from inverse Compton scattering of low-energy 
photons by the same relativistic electrons in the jet. Details of this 
model remain unresolved (e.g. see Hartman et al. 1997, for a review). 
The relative power output in $\gamma$-rays for B2013+370 is less than that of a 
typical flat-spectrum radio quasar seen by EGRET, and is more similar to that 
observed in BL Lac objects. 
The location of the broad synchrotron peak in the optical-IR band rather 
than in the X-ray band is an indication that B2013+370 could be a 
low-energy peaked BL Lac object (LBL), according to the classification 
suggested by Giommi \& Padovani (1994). 

We believe that the association of 3EG J2016+3657 with B2013+370 is 
real and that the source of both is most likely a blazar. 
\bigskip

In conclusion, we have made a comprehensive study of the X-ray and $\gamma$-ray 
sources in the error circle of the COS--B source 2CG 075+00. We have identified 
most of the X-ray sources in the error boxes of the EGRET sources. 
One reasonable hypothesis is that 2CG 075+00 is largely the same source as 
3EG J2021+3716 and that 3EG J2016+3657 is a new 
source which can be identified with the radio
blazar 2013+3710 = X-ray source 3. Herein 
2CG 075+00 = 3EG J2021+3716 remains unidentified.
Clearly, for the 3EG sources
considered here, we need more refined $\gamma$-ray positions and
extensive monitoring (possibly by AGILE and GLAST) 
to establish their ultimate nature. 

\acknowledgements

We thank Jonathan Kemp for obtaining optical images at MDM Observatory. 
We acknowledge support by NASA Grant NAG5-3696 (R. M.), NASA Grant 
NAG5--7935 (E. V. G.), and NASA Grant NAG5-3229 (J. P. H). 
This research has made
use of data obtained from HEASARC at Goddard Space Flight Center and 
the SIMBAD astronomical database.

\newpage

{\bf Figure Captions}

Fig. 1. --- ROSAT soft X-ray image of 3EG J2016+3657 and 3EG J2021+3716. The 
circles for the 
two 3EG sources correspond to the $\sim 95$ \% confidence contours. The dashed 
circle corresponds to the COS--B source 2CG 075+00. The GeV Catalog source 
is also shown. 

\medskip

Fig. 2. --- EGRET $\gamma$-ray light curves for (a) 3EG J2016+3657 and (b) 3EG J2021+3716 
from 1991 to 1997. $2 \sigma$ 
upper limits are shown as downward arrows. The horizontal error bars 
correspond to the extent of an individual observation. 

\medskip

Fig. 3. --- ASCA image of 3EG J2016+3657 and 3EG J2021+3716. The circles for the two 3EG sources 
correspond to the $\sim 95$ \% confidence contours. The dashed 
circle corresponds to the COS--B source 2CG 075+00. The GeV Catalog source 
is also shown. 

\medskip

Fig. 4. --- ROSAT HRI X-ray image of the field around 3EG J2016+3657. The image shows the sources
2 and 3 (B2013+370) as clearly resolved point sources. The image is scaled 
to highlight point features. 
The source numbers correspond to those listed in Table 1.

\medskip

Fig. 5. --- Flux history of B2013+370 at 90 GHz and 140 GHz between 1993 and 1995. 
(Data compiled from Reuter et al. 1997).

\medskip

Fig. 6. --- A section of the combined $R$-band CCD images from the MDM 2.4m 
telescope. Total exposure time was 30 min. 
The field shown is $70^{\prime\prime}\times 70^{\prime\prime}$.
Tick marks are centered on the NVSS position of the blazar
B2013+370. The optical object between the tick marks has
$R = 21.6 \pm 0.2$, and is located at (J2000)
20 15 28.76, +37 10 59.9.

\medskip

Fig. 7. --- ROSAT PSPC Image of 3EG J2016+3657 and 3EG J2021+3716 showing the 
bright 1.4 GHz radio sources from the NRAO/VLA Sky Survey 
within the error circles of the EGRET sources. Only 
those sources with radio fluxes $> 0.3$ Jy are shown. Refer to Table 2 for 
fluxes and source positions. The error circles of 2CG 075+00, the two EGRET 
catalog sources, and the GeV Catalog source are indicated. 

\medskip

Fig. 8. --- 
Broad band spectrum of B2013+370, assuming that it is the $\gamma$-ray 
source 3EG J2016+3657. 

\newpage

\begin{table}[h!]\scriptsize
\tablenum{1}
\begin{center}
\caption{X-ray sources in the fields of 3EG J2016+3657 and 3EG J2021+3716}
\begin{tabular}{clccccccc}
\tableline
Number$^a$& Source Name& RA & Dec& Count Rate$^b$ & $HR^c$ & Sig$^d$ & Count Rate$^b$ & Sig$^d$ \\
          &            &    &    & (ASCA)         & (ASCA) & (ASCA)  &  (ROSAT)       & (ROSAT) \\
\tableline
1  & CTB87          & 20 16 11.5  & +37 11 19    & $24\pm 0.7$ & $-0.42$& 28 & $ 7.5  \pm1.6 $ & 4.3 \\
2  &                & 20 15 36.6  & +37 11 25    & $30\pm 0.7$ & \dots  & 42 & $8.8  \pm1.6 $ & 5.1 \\
3  & B2013+370       & 20 15 28.3  & +37 11 02    & $15\pm 0.4$ & \dots  & 30 & \dots          & \dots  \\
4  &                & 20 15 35.7  & +37 04 59    & $ 9\pm 0.6$ & 0.0    & 12 & $19.0  \pm2.1$ & 8.8 \\
5  &                & 20 15 14.9  & +36 59 22    & $3\pm 0.4$  & $-0.26$& 5.5& $11.5  \pm1.8$ & 6.1 \\
6  & HD228766       & 20 17 29.0  & +37 18 28    & \dots       & \dots  & \dots& $13.6  \pm1.8$ & 7.3 \\
7  & HD193077       & 20 16 59.8  & +37 25 26    & \dots       & \dots  & \dots& $25.6  \pm2.2$ & 11.1\\
8  & Bright Star    & 20 16 37.4  & +37 05 55    & \dots       & \dots  & \dots& $12.4  \pm1.8$ & 6.5 \\
9  & Bright Star    & 20 17 34.6  & +36 38 06    & \dots       & \dots  & \dots& $2.6 \pm 9.1$  & 2.4 \\
10 &                & 20 23 22.2  & +37 59 42    & \dots       & \dots  & \dots& $40.6\pm 3.8$  &10.6 \\
11 &                & 20 20 42.4  & +38 02 05    & \dots       & \dots  & \dots& $16.5\pm 3.0$  & 5.2 \\
12&CCDM J20215+3758A& 20 21 21.6  & +37 58 15    & \dots       & \dots  & \dots& $7.7 \pm 2.4$  & 3.0 \\
13 & HD229153       & 20 22 45.7  & +37 57 18    & \dots       & \dots  & \dots& $3.9 \pm 2.0$  & 2.0 \\
14 & Bright Star    & 20 19 43.7  & +37 35 49    & \dots       & \dots  & \dots& $11.4\pm 1.8$  & 4.4 \\
\tableline
\end{tabular}

$(a)$  Identifying number in ROSAT image (Fig. 1). 
$(b)$ {Background-subtracted source count rate 
$\times 10^{-3}$ s$^{-1}$ extracted from a $2^\prime$ diameter aperture, 
in the 2 -- 10 keV energy band for ASCA and 0.5 -- 2 keV energy 
band for ROSAT (PSPC).}
$(c)$ Hardness ratio (ASCA). For sources 2 and 3 
this is not determined due to severe cross contamination.
$(d)$ Significance computed using method of Gotthelf \& Kaspi (1998). 

\end{center}
\end{table}

\newpage

\begin{table}[h!]\small
\tablenum{2}
\begin{center}
\caption{Radio sources ($>0.3$ Jy) within the error circles of 3EG J2016+3657 and 3EG J2021+3716}
\begin{tabular}{cccccl}
\tableline
NVSS  1.4 GHz  &            &Integrated      & WENSS (326 MHz) & Integrated & Other \\
RA (2000)      & Dec (2000) & Flux (Jy)      & Name            & Flux (Jy)  & Names\\
\tableline
20 15 28.73 & +37 10 59.9  & $2.17\pm 0.07$  & WNB2013.6+3701  &   1.14 & B2013+370\\
20 15 53.68 & +37 11 30.2  & $1.02\pm 0.03$  &       -         &        &
Associated with CTB 87  \\
20 16 04.13 & +37 12 23.5  & $1.81\pm 0.06$  & WNB2014.1+3703  &   1.69 & CTB 87\\
20 16 05.82 & +37 14 05.5  & $1.17\pm 0.04$  &       -         &        &
Associated with CTB 87 \\
20 17 46.86 & +36 44 47.9  & $0.29\pm 0.01$  &       -         &        & H {\tt
II} region S104\\
            &              &                 &                 &        &   \\
20 21 38.67 & +37 31 10.1  & $6.58\pm 0.24$  & WNB2019.7+3718B &   1.49 & H {\tt II} region\\
20 21 41.60 & +37 25 52.1  & $2.98\pm 0.11$  & WNB2019.7+3718C &   1.98 & H {\tt II} region\\
\tableline
\end{tabular}
\end{center}
\end{table}
\newpage

\begin{figure}[h!] 
\centerline{\epsfig{file=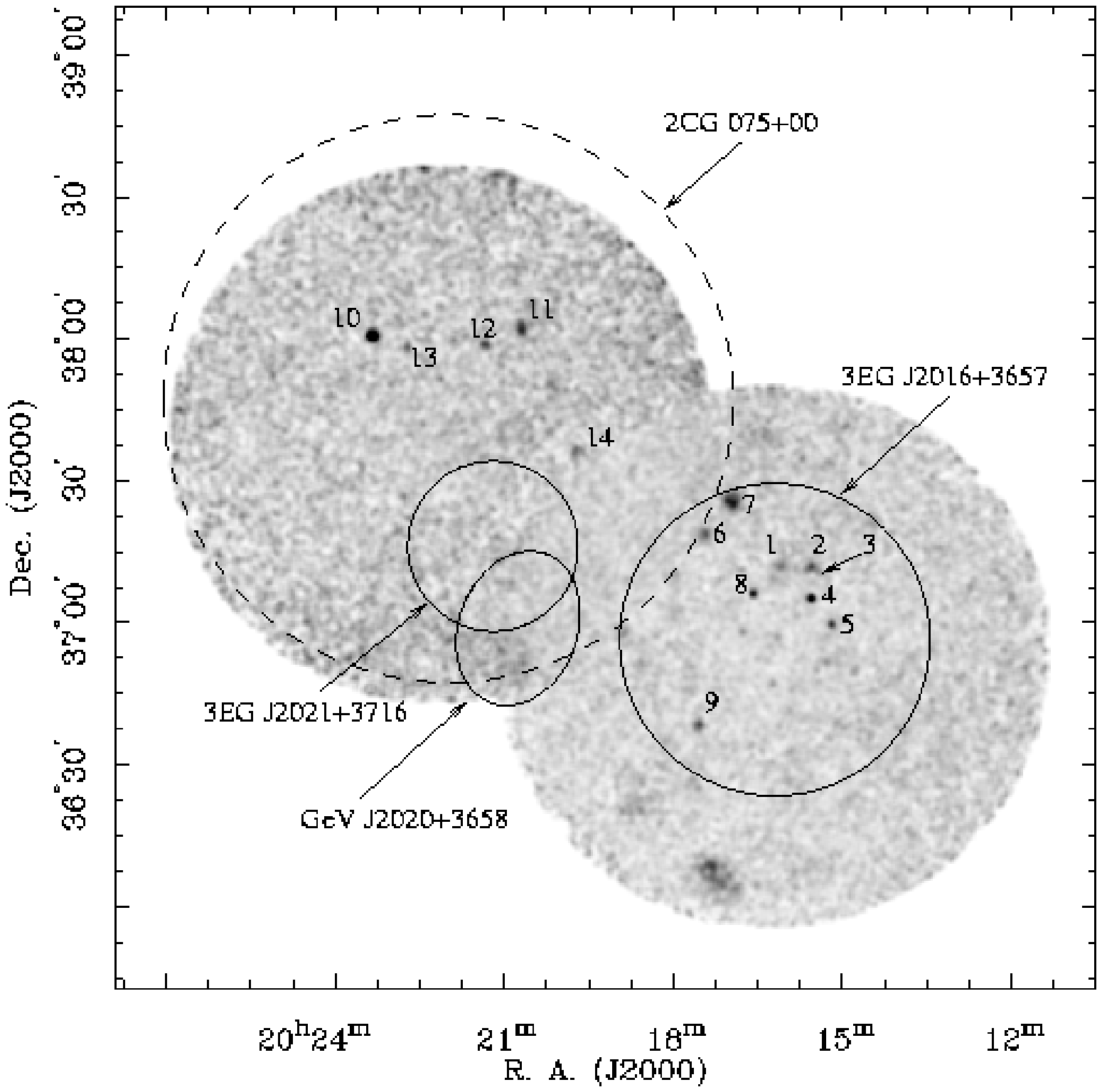,height=6.0in,bbllx=45pt,bblly=135pt,bburx=575pt,bbury=655pt,clip=.}}
\vspace{10pt}
\caption{ROSAT soft X-ray image of 
3EG J2016+3657 and 3EG J2021+3716. The circles for the 
two 3EG sources correspond to the $\sim 95$ \% confidence contours. The dashed 
circle corresponds to the COS--B source 2CG 075+00. The GeV Catalog source 
is also shown. }
\label{fig1}
\end{figure}

\newpage

\begin{figure}[t!] 
\centerline{\epsfig{file=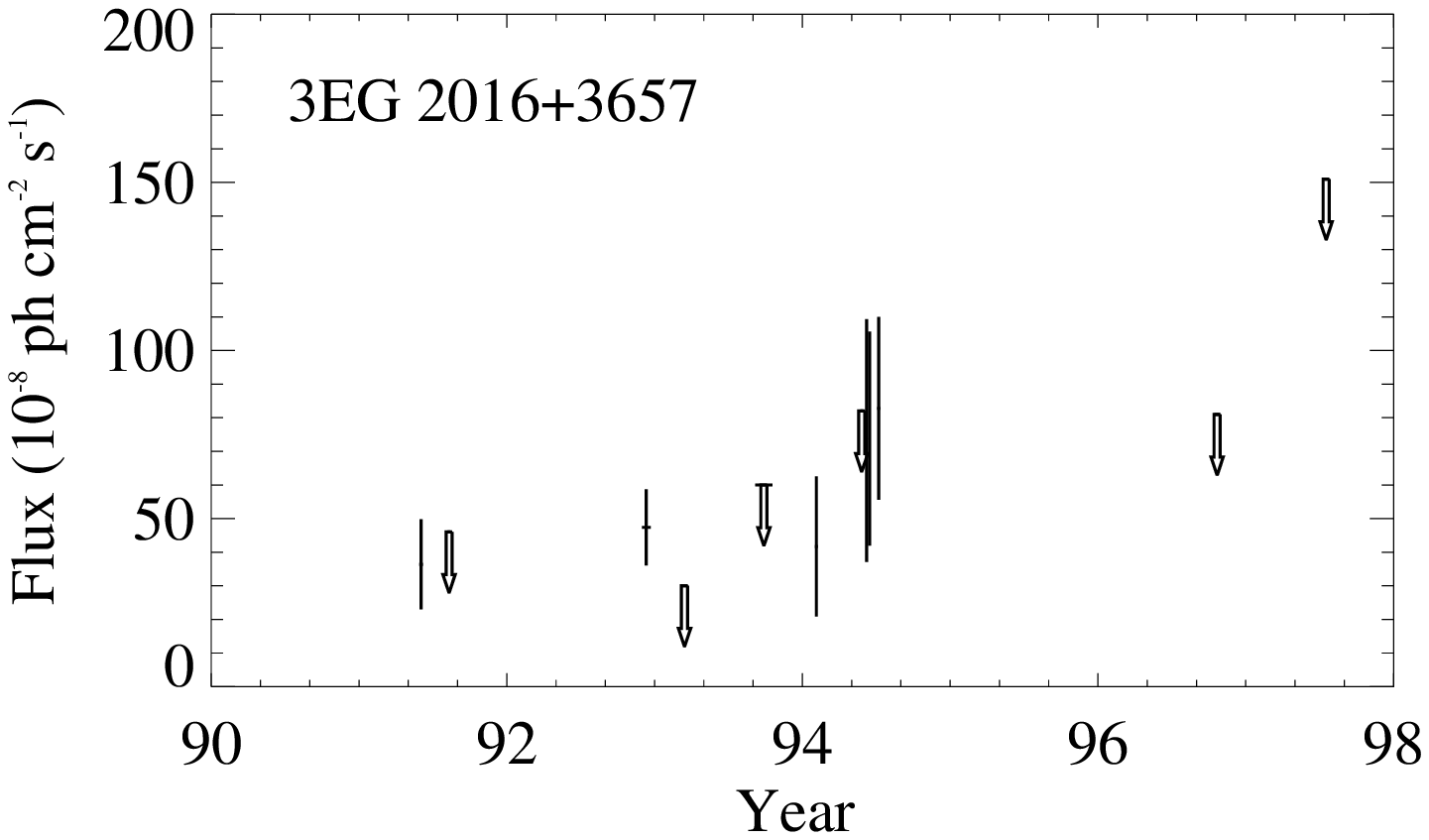,height=3.0in,bbllx=50pt,bblly=275pt,bburx=475pt,bbury=530pt,clip=.}}
\vspace{3pt}
\centerline{\epsfig{file=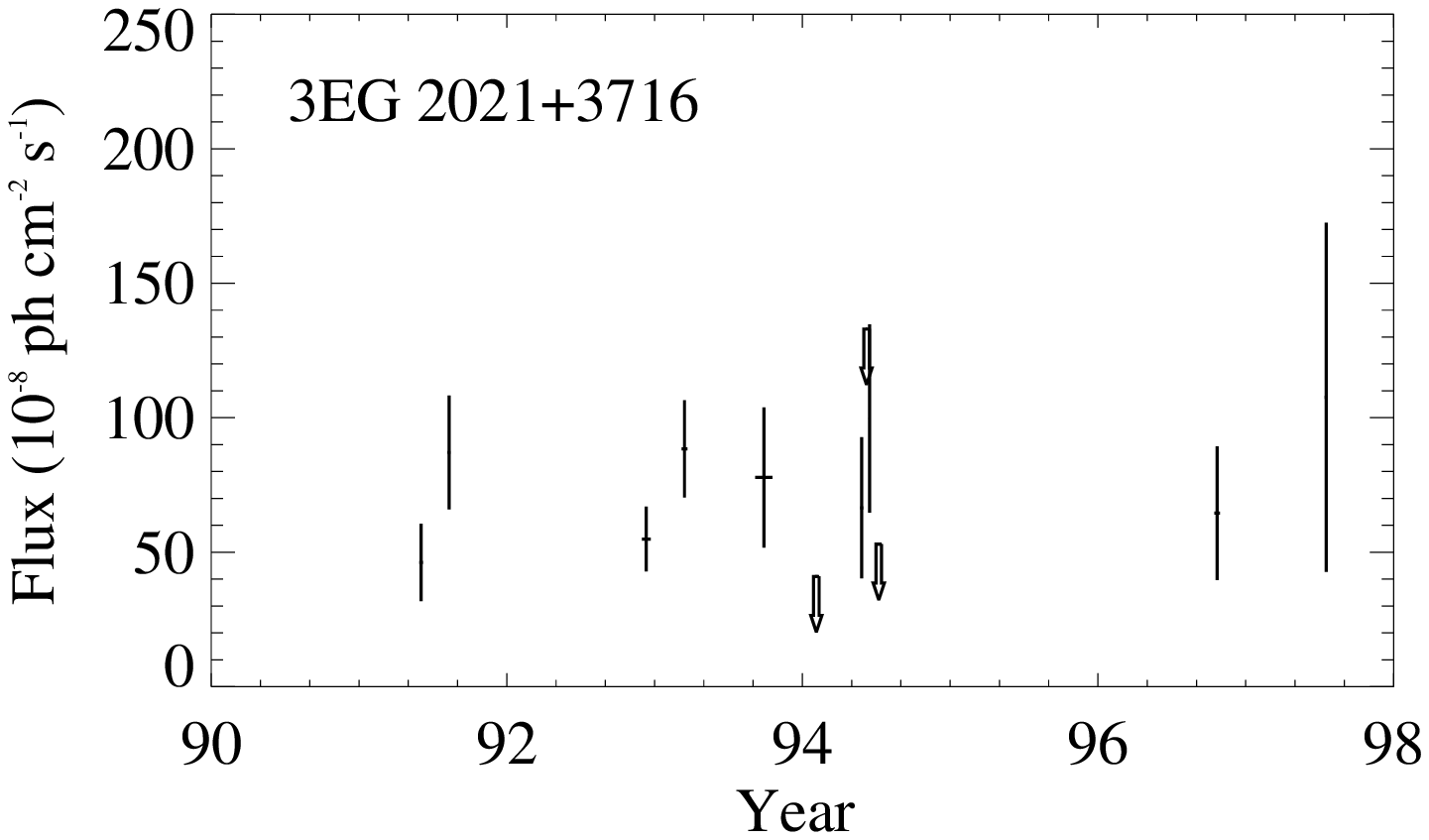,height=3.0in,bbllx=50pt,bblly=275pt,bburx=475pt,bbury=530pt,clip=.}}
\vspace{10pt}
\caption{EGRET $\gamma$-ray light curves for (a) 3EG J2016+3657 and (b) 3EG J2021+3716 
from 1991 to 1997. $2 \sigma$ 
upper limits are shown as downward arrows. The horizontal error bars 
correspond to the extent of an individual observation.}
\label{fig2}
\end{figure}
 
\newpage

\begin{figure}[h!] 
\centerline{\epsfig{file=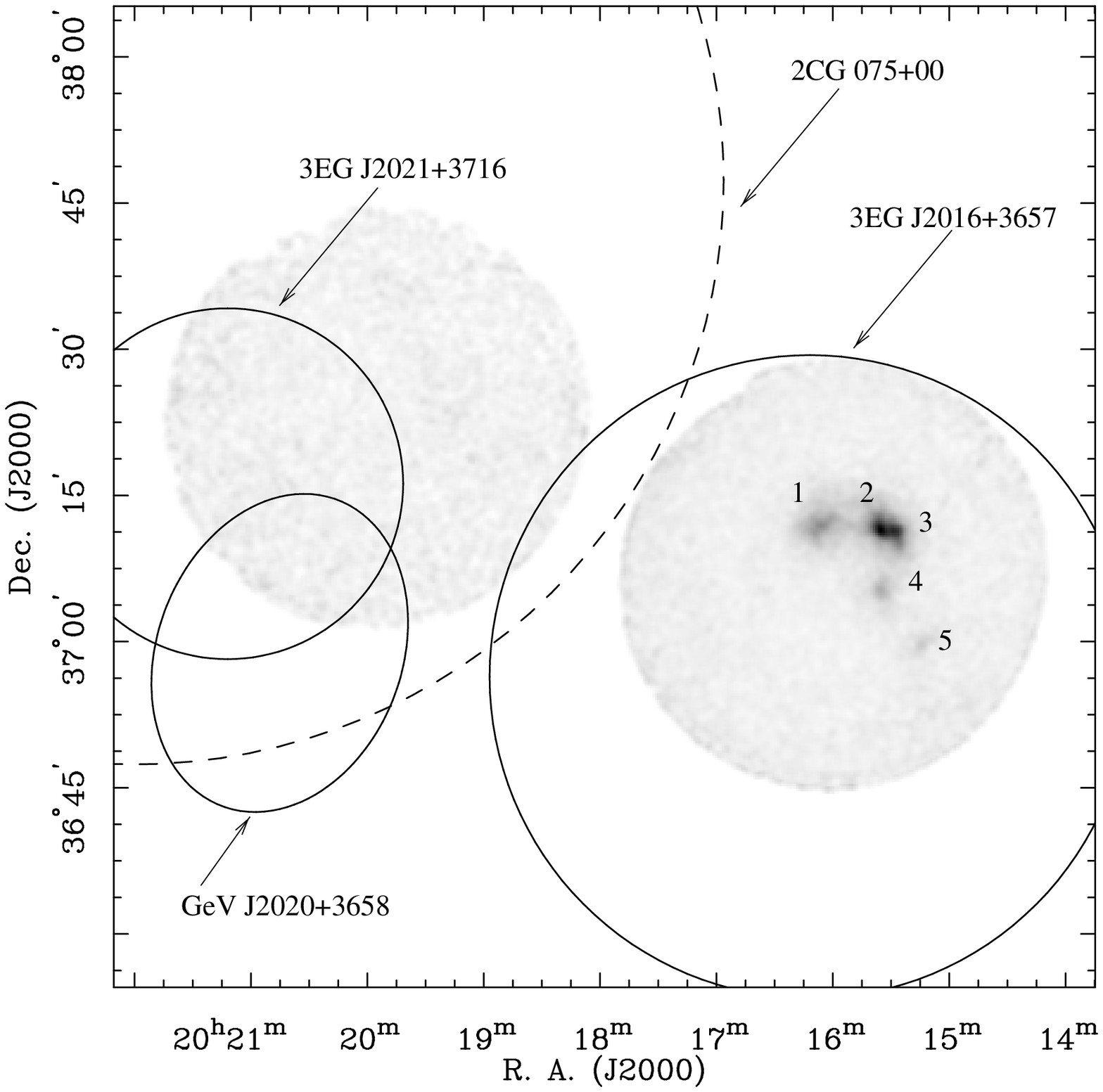,height=6.0in,bbllx=50pt,bblly=110pt,bburx=575pt,bbury=655pt,clip=.}}
\vspace{10pt}
\caption{ASCA image of 3EG J2016+3657 and 3EG J2021+3716. The circles for the two 3EG sources 
correspond to the $\sim 95$ \% confidence contours. The dashed 
circle corresponds to the COS--B source 2CG 075+00. The GeV Catalog source 
is also shown. }
\label{fig3}
\end{figure}

\newpage

\begin{figure}[h!] 
\centerline{\epsfig{file=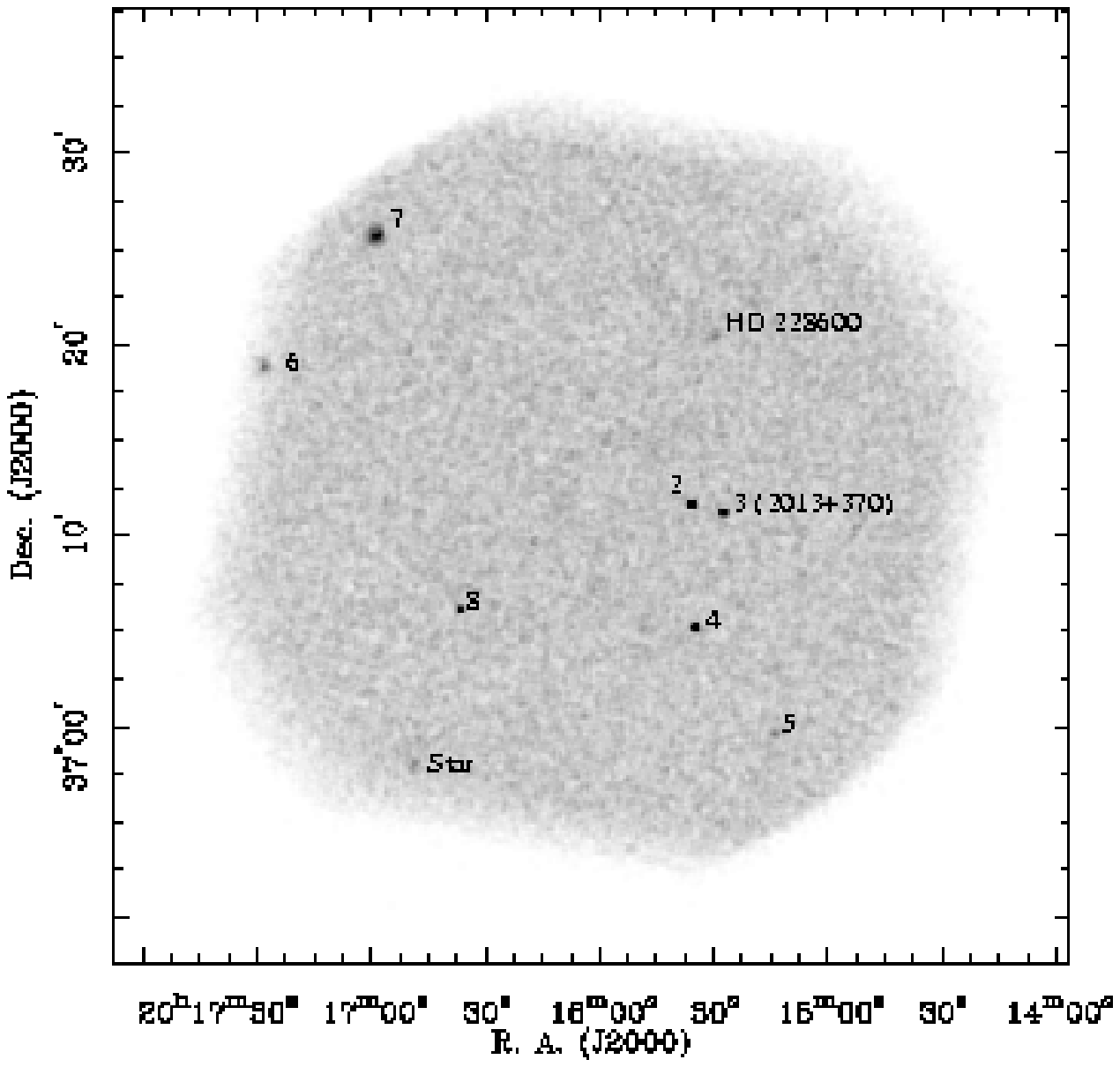,height=5.0in,bbllx=35pt,bblly=140pt,bburx=580pt,bbury=660pt,clip=.}}
\vspace{10pt}
\caption{ROSAT HRI X-ray image of the field around 3EG J2016+3657. The image shows the sources
2 and 3 (B2013+370) as clearly resolved point sources. The image is scaled to 
highlight point features. 
The source numbers correspond to those listed in Table 1.}
\label{fig4}
\end{figure}
 
\newpage

\begin{figure}[h!] 
\centerline{\epsfig{file=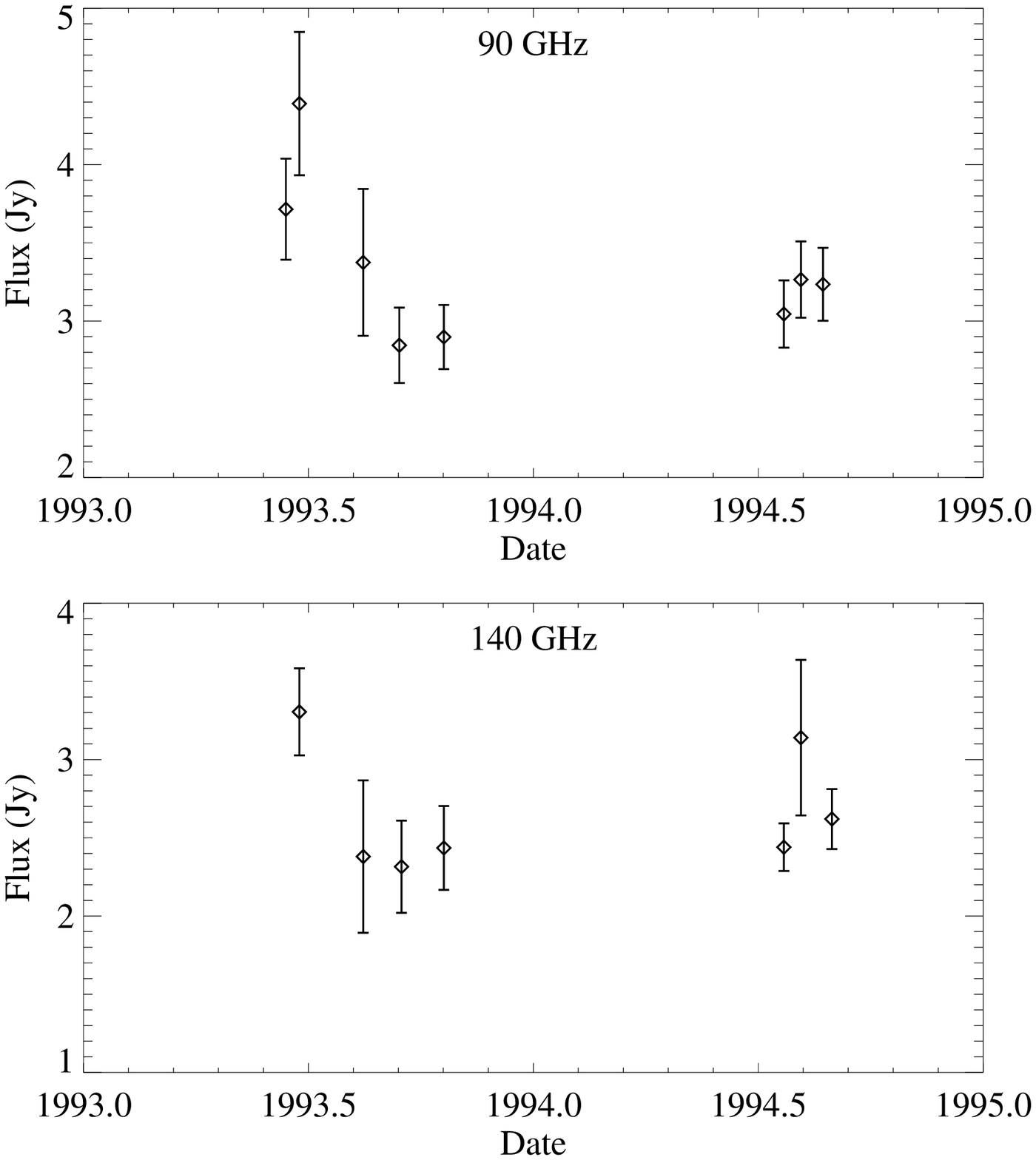,height=5.0in,bbllx=55pt,bblly=25pt,bburx=565pt,bbury=595pt,clip=.}}
\vspace{10pt}
\caption{Flux history of B2013+370 at 90 GHz and 140 GHz between 1993 and 1995. 
(Data compiled from Reuter et al. 1997).} 
\label{fig5}
\end{figure}
 
\newpage

\begin{figure}[h!] 
\centerline{\epsfig{file=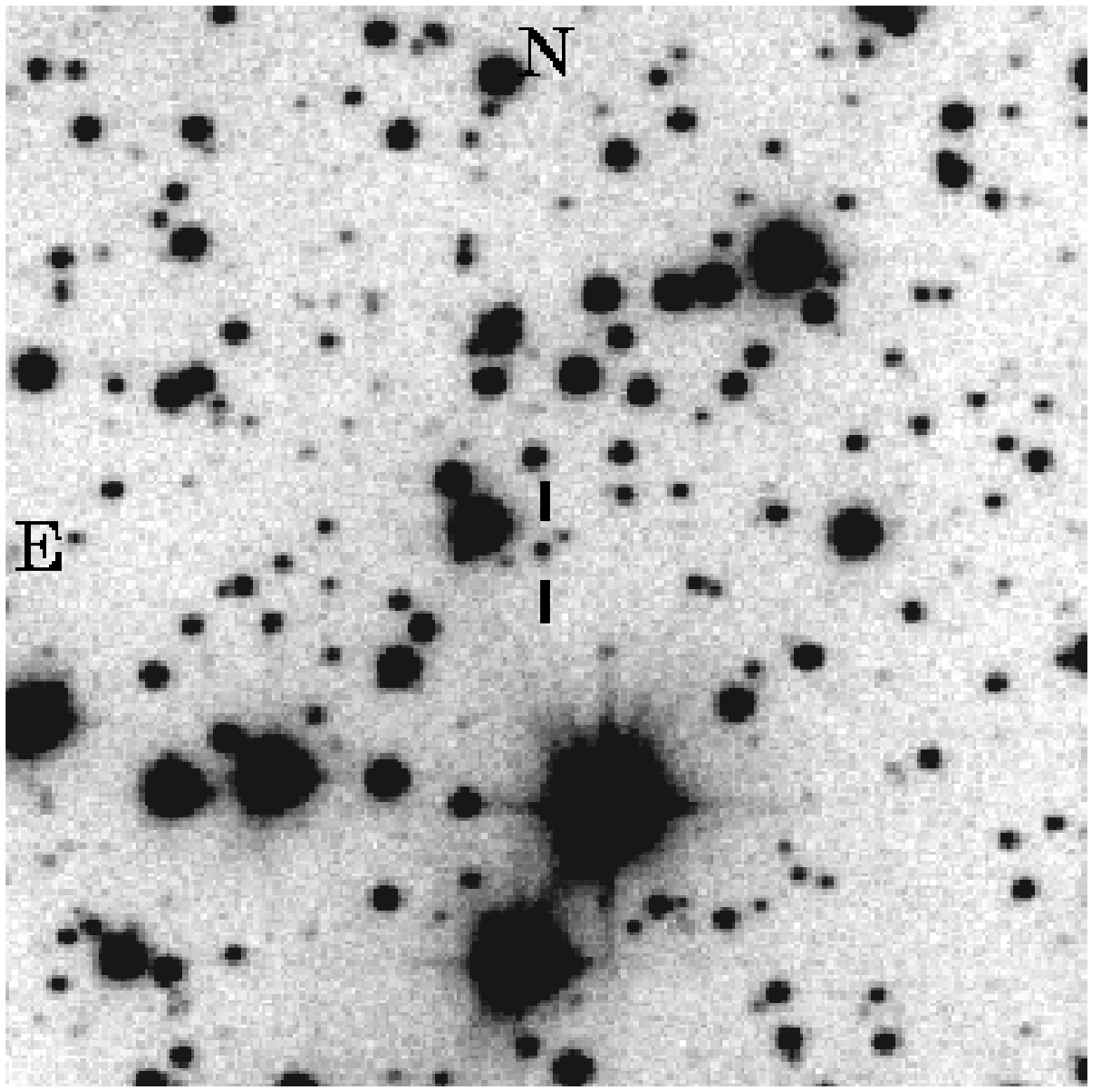,height=5.0in,bbllx=50pt,bblly=140pt,bburx=570pt,bbury=650pt,clip=.}}
\vspace{10pt}
\caption{A section of the combined $R$-band CCD images from the MDM 2.4m 
telescope. Total exposure time was 30 min. 
The field shown is $70^{\prime\prime}\times 70^{\prime\prime}$.
Tick marks are centered on the NVSS position of the blazar
B2013+370. The optical object between the tick marks has
$R = 21.6 \pm 0.2$, and is located at (J2000)
20 15 28.76, +37 10 59.9.}

\label{fig6}
\end{figure}
 
\newpage

\begin{figure}[h!] 
\centerline{\epsfig{file=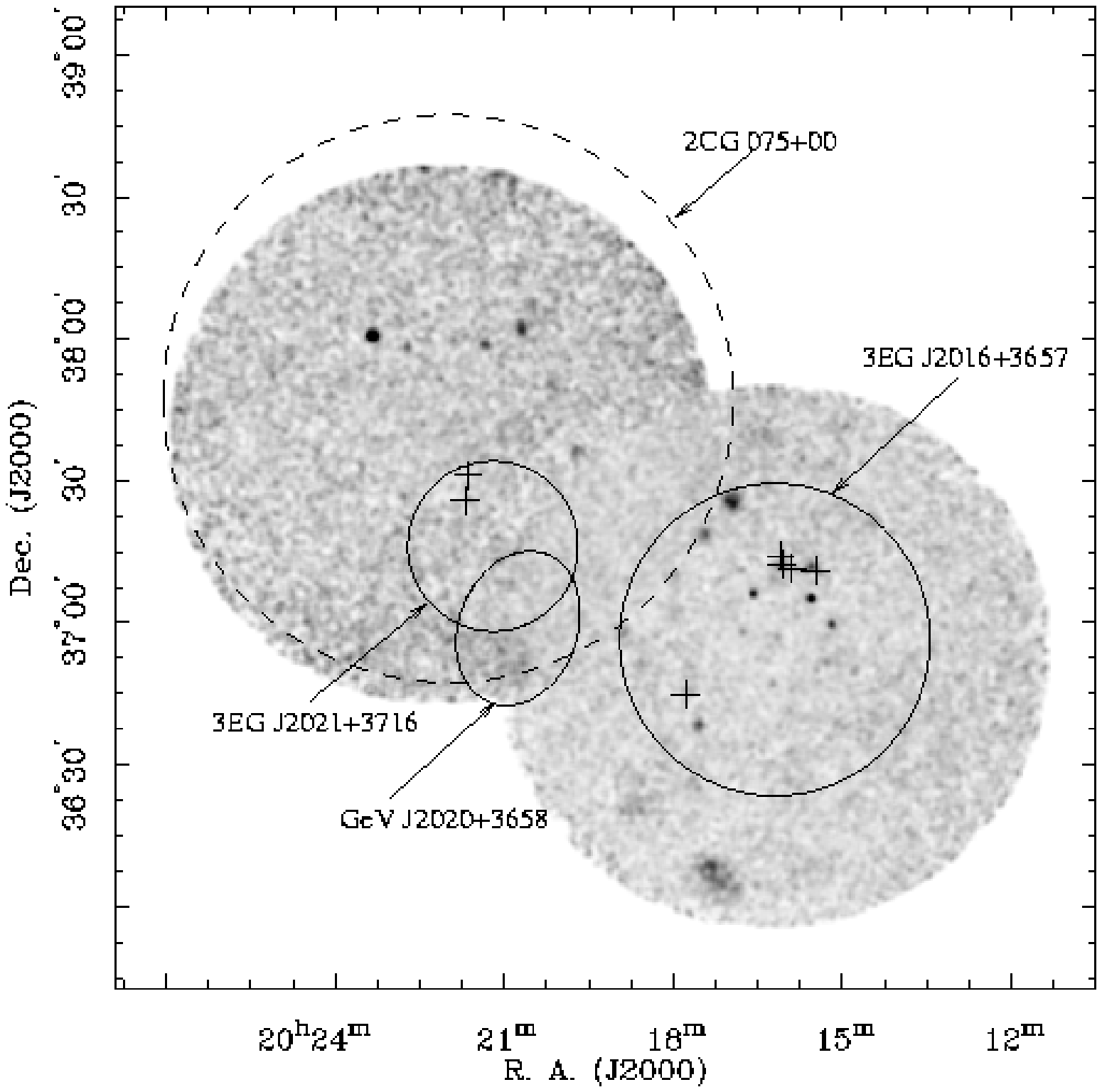,height=6.0in,bbllx=50pt,bblly=110pt,bburx=575pt,bbury=655pt,clip=.}}
\vspace{10pt}
\caption{ROSAT PSPC Image of 3EG J2016+3657 and 3EG J2021+3716 showing the 
bright 1.4 GHz radio sources from the NRAO/VLA Sky Survey 
within the error circles of the EGRET sources. Only 
those sources with radio fluxes $> 0.3$ Jy are shown. Refer to Table 2 for 
fluxes and source positions. The error circles of 2CG 075+00, the two EGRET 
catalog sources, and the GeV Catalog source are indicated. }
\label{fig7}
\end{figure}
 
\newpage

\begin{figure}[b!] 
\centerline{\epsfig{file=muk_fig8.ps.gz,height=4.0in,bbllx=0pt,bblly=50pt,bburx=590pt,bbury=400pt,clip=.}}
\vspace{10pt}
\caption{Broad band spectrum of B2013+370, assuming that it is the $\gamma$-ray 
source 3EG J2016+3657. }
\label{fig8}
\end{figure}

\end{document}